# Plasma assisted CO₂ splitting to carbon and oxygen: a concept review analysis.

G. Centi[*] and S. Perathoner

[a] *Dept. ChiBioFarAm, University of Messina, V-le F. Stagno D'Alcontres 31, 98166 Messina (Italy), and ERIC aisbl, Bruxelles (Belgium).*

## A B S T R A C T

This concept review paper analyses the possibility to develop a technical solution for the challenging reaction of CO₂ splitting to carbon and O₂ (CO₂tC), a dream reaction in the area of addressing climate change and greenhouse gas emissions. There are quite limited literature indications that were reviewed, with an analysis also of the limits. Not-thermal plasma, in combination with catalysis, is one of the possibilities to promote this reaction, but the current studies on CO₂ splitting are limited mainly to the formation of CO. By combining data on the status of current studies on plasma CO₂ splitting with those on CO₂tC it is possible to propose a novel approach which is presented and discussed and aims to stimulate creativity in consider alternative possibilities for this challenging reaction, rather than to demonstrate feasibility of the proposed technology.

## 1. Introduction

The topic of $CO_2$ conversion and utilization is among those with the largest research interest in the recent years, with over 250 reviews in the last three years centred on various aspects, from thermochemical catalytic approaches (the various Power-to-X technologies [1,2], or catalytic/biocatalytic routes to produce valuable chemicals and monomers [3-6]) to photo and electro-catalytic approaches [7-13] and plasma-catalysis [14-19]. Notwithstanding this very large research interest, there are extremely few results in literature on what can be considered a dream reaction to reduce $CO_2$ emissions, the carbon dioxide splitting (also called cracking, dissociation or decomposition, eq. 1) which is the equivalent of the widely investigated methane splitting (or cracking/decomposition/pyrolysis, eq. 2 [20-22]):

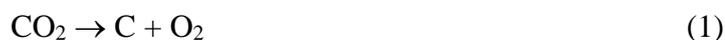

$$CO_2 \rightarrow C + O_2 \qquad (1)$$

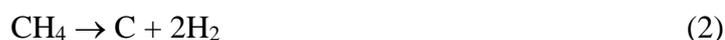

$$CH_4 \rightarrow C + 2H_2 \qquad (2)$$

In terms of sustainability $CO_2$ splitting to C + $O_2$ is preferable over the reaction with $H_2$ (or H-donors) because it avoids the energy necessary for water splitting (or production of $H_2$ or H-donors), thus resulting potentially more efficient from the energy use.

$CO_2$ splitting essentially reverses coal combustion (carbon plus oxygen yields $CO_2$ and energy), and by supplying renewable energy for the transformation, could become a very important complementary process to close the carbon cycle, particularly in energy intensive industries. It is a better alternative to $CO_2$ sequestration and storage (CCS), releasing back oxygen and forming carbon which can more safely and less energy-intensive stored with respect to $CO_2$. In addition, the produced carbon, can be used as carbonaceous materials depending on which form is generated during the splitting process (for example, carbon nanotubes, graphene-like species, amorphous carbon, etc.).

The utilization of advanced carbon materials is increasing worldwide (global market for advanced carbon materials is worthy for over USD 7 billion and is in further fast expansion) or the produced carbon from $CO_2$ dissociation can be utilized as soil amendment (biochar). Biochair use decreases GHG (greenhouse gas) emissions and acts as long-term carbon sequestration medium. For this reason, biochar is considered a carbon negative solution [23,24]. With respect to the routes of $CO_2$ utilization, $CO_2$ splitting technology has the potential to overcome the limitations of many of the explored routes related to market size of the products and can have a larger impact in terms of reduction of GHG emissions, being one of the few carbon-negative technologies. Furthermore, the potential range of application is wide, from distributed productions (converting $CO_2$ in biogas) to large industrial cases ($CO_2$ emissions in steel and iron manufacture, cement, refineries etc.).

In front of these potential advantages, the studies on this reaction are very limited, and often when $CO_2$ splitting or decomposition is indicated, the reaction of carbon dioxide conversion to CO is instead reported. This reaction to form CO rather than C has been extensively studied by different approaches, from thermochemical (solar) cycles to thermocatalytic conversion (for example, reverse water gas shift or dry reforming), and by using electro-, photo- and plasma-catalysis processes, although typically in the presence of an H-donor [25-30]. The further CO conversion to carbon by direct CO decomposition into C + O occurs only when these species may be further converted, but the catalytic disproportionation, e.g. the Boudouard reaction - eq. 3 [31]) is well known:

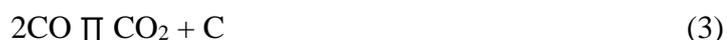

$$2CO \rightleftharpoons CO_2 + C \qquad (3)$$

The temperature is an important element determining the equilibrium of this reaction. At temperatures about above 700°C the reaction in shifted to the left, i.e. to the formation of carbon monoxide. This equilibrium reaction competes with eq. 1, limiting the possibility to maximize the $CO_2$ splitting. CO decomposition at the surface of the catalysts is often an important elementary step in catalytic mechanisms of $CO_x$ conversion, such as the formation of hydrocarbons in the Fischer Tropsch conversion of syngas [32], but these surface-bound C species deriving from the dissociation are fast hydrogenated and further converted, thus avoiding the blocking of the surface sites. In general, when carbon forms in the process, it tends to fast deactivate the surface [33]. Carbon forms also in eq. 2, but also in reforming reactions, segregates forming species (like carbon filaments) [34] avoiding a fast deactivation of the catalyst surface [35]. Oxygen also tends to remain on the catalyst surface blocking the reactivity [36]. Addition of H-donors (methane, as in dry reforming) remove both these species (surface-bound O and C) and maintains catalyst activity.

Implementing thus the $CO_2$ splitting process (eq. 1) would require i) to operate at temperatures not too high to shift equilibrium in eq. 1 in favour of CO formation, while at the same time allowing enough high kinetics of reaction to make industrially feasible the process, and ii) produce a C species avoiding a fast deactivation of the catalysts. It is also necessary that the C species (carbon filaments or other carbon species) could be removed from the catalyst to allow a continuous process and be in a marketable form. Meet these requirements is the great challenge of $CO_2$ splitting to C and $O_2$ explaining also why still very limited progresses have been made in this direction.

Non-thermal plasma (NTP) is a technology of increasing relevance also in the field of $CO_2$ conversion [14-18, 29, 37-41], but the studies were centred nearly exclusively on the reactions of i) $CO_2$ splitting to CO and $O_2$ (rather than to C and $O_2$), ii) $CO_2$ hydrogenation or iii) reforming in the presence of $H_2O$ and/or $CH_4$ and the interaction between the species generated in gas phase by NTP and the eventual catalyst [37]. Therefore, NTP even in the presence of solid catalysts is not able itself to effectively promote the $CO_2$ to C + $O_2$ splitting (eq. 1) but needs to be integrated with other approaches to attempt to develop a technology for this challenging reaction.

This contribution is indicated with the term of "concept review analysis" because aims not only to analyse the state of the art in $CO_2$ splitting to C + $O_2$ as the basis to identify new possibilities of research directions to address this reaction. It is organized in a first part dealing on the limited studies and attempts to make this challenging reaction, a second part of brief analysis of the state of the art in $CO_2$ splitting by NTP (without the presence of possible H-donor), and a final part of analysis of all the results to identify possible new research directions to progress in this reaction of $CO_2$ splitting to C and oxygen.

## 2. Status on $CO_2$ splitting to carbon and oxygen (CO₂tC)

The studies in literature on $CO_2$ splitting to carbon and oxygen (CO₂tC) are quite limited. One of the more specific studies is that made by Luc et al. [42] investigating $CO_2$ splitting using an electro-thermochemical hybrid looping (ETHL) strategy. The ETHL process combines two steps. A first electrochemical conversion of $CO_2$ to CO which couples with a thermochemical further conversion of CO into C and $CO_2$. The latter product ($CO_2$) is recirculated to the first step. In this way, it is possible to obtain 96% selectivity in the first $CO_2$ electrolysis step and nearly 100% selectivity for the CO thermochemical step. The schematic representation of the ETHL process is shown in Figure 1. Luc et al. [42] proved the concept of the single steps, but not of the fully integrated device. For the $CO_2$ electrolyzer they used a 25 cm$^2$ gold-plated stainless steel continuous flow-cell type electrolyzer where Ag catalyst was used to facilitate the conversion of $CO_2$ to CO while Ir nanoparticles (deposited on a Nafion XL membrane) were used for the water oxidation to $O_2$. Thus, although apparently the process converts $CO_2$ to C + $O_2$, an in-situ water oxidation at the anode forming $O_2$ and protons is present. The $H^+$ then are used to hydrogenate $CO_2$ to form CO and water, the latter remaining in the electrolyte. In considering the energy necessary for the process, that required for the water electrolysis step ($\Delta G_{298}$ = -457.2 kJ) must be considered. Thus, even if the $\Delta G_{298}$ for $CO_2$ to C + $O_2$ is 394.4 kJ and that for $CO_2$ + 2 $H_2$ → C + 2$H_2O$ is -62.8 kJ, the energy required for the latter process is higher than that of the direct splitting $CO_2 \rightarrow C + O_2$ (even higher considering the overpotential required in the electrolysis). Thus, to realize the direct splitting would be a less endoergic process.

The anode part of the $CO_2$ electrolyzer is in direct contact with air, while the cathodic part operates in contact with a flow of $CO_2$-saturated 0.5 M $NaHCO_3$ aqueous electrolyte solution. Although other type of $CO_2$ electrolyzers /electrodes could be also used, the one used by Luc et al. [42] is efficient with formation of only CO (except traces of formate), a good total current density (around 30 mA·cm$^{-2}$) and the conditions of maximum CO Faradaic efficiency - FE (96%) reached at 2.8 V of applied potential. However, in about 2h the $FE_{CO}$ decreases to about 70% and the current density nearly halves, attributed to $Na^+/H^+$ cation exchange within the Nafion membrane with decrease of ionic conductivity and increase in internal resistance within the electrolyzer. By regenerating the membrane with dilute sulfuric acid, the performance of the $CO_2$ electrolyzer are recovered. This is not a practical solution for a process, but more stable $CO_2$ electrolyzers could be developed, which also do not need to operate in contact with sodium bicarbonate electrolytes. So called gas-phase $CO_2$ electrolyzers, which do not have a bulk liquid electrolyte, but rather use gas-diffusion layer (GDL) type electrodes in direct contact with the Nafion membrane, can be used [43], or alternatively ionic liquids can be used as electrolytes [44]. Thus, the problem of stability can be overcome in principle. Still remain the problem to analyse the long-term stability, and the presence of other components in the inlet $CO_2$ stream to the electrolyzers, because during continuous operations, the byproducts formed will accumulate (even considering a purge) and CO will be also present, because it will be not fully converted in the CO splitting unit.

For the CO thermochemical conversion, based on the Boudouard reaction (eq. 3), Luc et al. [42] used a tubular reactor assembly operating in a semi-continuous fashion with the removal/replenish of the catalyst pellet made routinely to remove the carbon deposits. The latter is then eliminated by thermal oxidation. This is not a suitable method for a project of $CO_2$ splitting, because $CO_2$ is reformed during the regeneration. Thus, it is necessary to optimize the type of carbon formed [45], by working on both the operation conditions and the type of catalysts, and in relation to the efficient removal of the carbon from the catalyst by a method allowing the carbon recovery.

There are very limited studies on the recovery of carbon deposits from spent catalysts, being most of the studies focused on the oxidative thermal regeneration (gasification) by using steam, air/oxygen, or $CO_2$. Parmar et al. [46] recently reported the possibility to extract the carbon (deposited in the form of carbon nanotubes - CNTs - during methane decomposition in a fluidized bed reactor) by a solvent-based ultrasonication method. The efficiency of carbon recovery depends strongly on the characteristics and amount of the carbon deposited, which in turn depends on the catalyst characteristics and operation conditions for methane decomposition.

Thus, there is space for improvement of the method, although the only focus in literature on gasification procedures (which even for methane splitting, eq. 2, is not a well-suited procedure, if the target is a $CO_x$-free $H_2$ production) should be remarked. Even when the method is patented to produce for example graphene (using iron ore catalysts) [47] there are no indications how the carbon material can then be effectively recovered.

Notwithstanding these limited literature indications on how to recover the carbon deposited on the CO splitting catalysts, by combining a proper selection of catalysts and operative conditions, and procedure of carbon extraction and catalyst regeneration, it is possible to consider feasible (in principle) the establishment of a continuous process of carbon extraction and CO splitting catalyst regeneration, to enable to couple with the $CO_2$ electrolysis unit and operating in a continuous manner the process outlined in Figure 1 (electro-thermochemical hybrid looping - ETHL). Note that even if apparently the process converts $CO_2$ to C + O2, the process involves the conversion of $CO_2$ to C + $H_2O$, plus the generation of $O_2$ at the anode side (by water oxidation). Thus, in estimating the energy required by the process, the water oxidation should be also accounted.

Luc et al. [42] used steel wool as CO splitting catalyst. The catalyst must be pretreated with acetic acid, and without this pretreatment, carbon deposition results much more limited. This appears a limitation for a continuous process, but there are many alternative catalysts active in the Boudouard reaction, which were not explored by Luc et al. [42] limiting their analysis to Fe and $Fe_2O_3$, and to 500°C as operational temperature, using as model for the study the direct carbon solid oxide fuel cells, but which typically operate at higher temperatures [48-50]. Note that the Boudouard reaction can be significantly promoted using microwaves [31]. In the conversion of $CO_2$ to CO and semi-coke (a "blue" coke that can replace metallurgical coke) Dai et al. [31] reported a quite large activity enhancement by microwave heating. Under microwave irradiation, dramatically different thermodynamics for the reaction are observed [51]. The apparent activation energy and the enthalpy of the reaction dropped largely under microwave irradiation. In addition, growing of specific nanocarbon structures can be made by microwave-assisted fabrication [52-54]. Therefore, operations in presence of microwave irradiation and NTP can add a further dimension to control the formation of carbon materials during the Boudouard reaction in terms of operation conditions and type of carbon formed, with related implications also on their possibility of selective removal from the CO splitting catalysts to enable continuous operations. These studies, however, not analysed how to make a continuous process, both in terms of removal of carbon for its utilization and recycle of the substrate.

Although Luc et al. [42] results are preliminary and several relevant aspects to develop a process are missing, the analysis made above indicates some valuable directions to improve the process, even if a general lack of literature knowledge on these aspects should be remarked. Note also that this $CO_2$toC reaction is also of high interest for missions beyond Earth's orbit, to realize an efficient oxygen recovery. A direct splitting of $CO_2$ requires extremely high reaction temperatures ($\gg$2000 K, $\Delta G^o$ = 394.4 kJ·mole$^{-1}$; [55]) which are not suitable on spacecrafts, but a process operating under mild conditions would be of high interest.

Other specific studies on $CO_2$tC reactions are not present in literature, but there are some data in support of some aspects discussed above. Kodama et al. [56] studied several years ago the $CO_2$ decomposition reaction into carbon at 300 °C, but using a $H_2$-prereduced Zn(II)-bearing ferrite (ZnO and wüstite, e.g. the latter being the precursor of ammonia synthesis catalyst). They also indicated that the decomposition reaction of $CO_2$ proceeds in two steps: (1) $CO_2$ reduction to CO, and (2) CO decomposition into carbon. The study, however, was not finalized to develop a $CO_2$tC process, but to investigate the chemistry of transformation of this catalyst during the processes of activation to produce the ammonia catalyst. A somewhat similar study was made by Zhang et al. [57] who studied the interaction of $CO_2$ with prereduced magnetite, other precursor for ammonia synthesis catalysts. $CO_2$ reacts with $Fe_3O_{4-\delta}$ also in a two-step process ($CO_2 \rightarrow CO + O^{2-}$ and $CO \rightarrow C + O$, the former being the slower process), forming $Fe_3O_4$ and carbon, although not indicating in which form the latter is generated. The process is not catalytic, but a reoxidation of the reduced iron-oxide by $CO_2$ with side formation of carbon occurs. They also indicated (but have not proven) that the carbon can be removed by ultrasonic treatment in a solvent, although mentioning that when a carbide species ($Fe_3C$) forms, this carbon removal procedure is not effective.

It is also known that graphene or graphene-like materials could be produced by $CO_2$ decomposition. Sun and Hu [58] reviewed the topic recently. There are different methods to produce graphene-like materials from $CO_2$: i) by electrolysis [59-61], ii) chemical vapor deposition - CVD [62] and iii) by alkali-metal chemistry [63-67], the latter being the method discussed in a more detail by Sun and Hu [58]. These and other studies on the production of graphene-like materials from $CO_2$ decomposition demonstrate the possibility to obtain controlled 3D porous graphene materials with intriguing properties for use in solar cells and as energy materials, but the studies were finalized to obtain these materials, not to develop a catalytic $CO_2$tC process. In fact, those described are largely stoichiometric processes, either induced by a prereduction of the material, as discussed before, or by a reaction such as the following in the alkali-metal chemistry [66]:

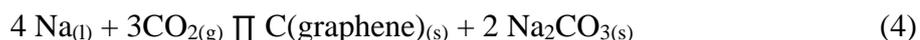

$$4\ Na_{(l)} + 3CO_{2(g)} \prod C(graphene)_{(s)} + 2\ Na_2CO_{3(s)} \qquad (4)$$

where the pedix (l) or (g) indicates whether is in liquid or gas form. This process allows to obtain 3D surface-microporous graphene due to this sequence of steps [68]: 1) $CO_2$ reacts (at around 520°C) with $Na_{(l)}$ to generate graphene sheets and $Na_2CO_3$ nanoparticles, the latter preventing the graphene sheets to aggregate, (2) the graphene walls can be oxidized by $CO_2$ forming surface micropores, and (3) the $Na_2CO_3$ is finally removed from the solid obtained using an aqueous solution of HCl leaving the 3D graphene-like material. Thus, a procedure not finalized to develop a continuous method of $CO_2$ decomposition to carbon and oxygen, but it shows the possibility to obtain valuable nanocarbon materials from the $CO_2$ splitting process. The CDV method is also finalized to produce the graphene material, although in this case there is in part also a catalytic role of the substrate (Cu-Pd deposited by sputtering 0001-sapphire chip [62]) in the reductive conversion of $CO_2$. A feed of $CO_2$ and $H_2$ is send to this material at 1000°C. After cooling, the graphene sheet is removed and transferred to a Si substrate forming on it a poly(methyl methacrylate) (PMMA) film, that allowed to transfer to a Si water, removing then the PMMA by chemical etching. As evident from the description, it is not a method which allows to develop a $CO_2$tC process.

The electrolysis method [59] is based on $CO_2$ absorption in a molten NaCl–CaCl electrolyte forming $O^{2-}$ and $CO_3^{2-}$, with the carbonate ion being then reduced at a cathode (stainless steel) at around 750°C in a two-step process to C+2 $O^{2-}$, the $O^{2-}$ being oxidized to $O_2$ at the anode side (RuO₂–TiO₂). CO also forms, but its amount not quantified. By changing the electrode (from stainless steel, to Cu, and Ni) or the temperature, the type of carbon material formed changes from graphite flake to graphene ball and few-layers flat graphene sheets. Also in this case, the possibility of continuous operations is not investigated.

Esrafilzadeh et al. [60] used instead a different approach, based on Ce-containing liquid metal (an euctectic Ga-In-Sn alloy was used as a liquid metal - LM) to make $CO_2$ electrolysis to carbon at room temperature. After adding cerium powder to LM, a drop of this liquid is deposited on a conductive carbon fibre to prepare the electrode. The droplet maintains its shape due to the high surface tension, but this procedure shows severe issues in scaling up the method. The role of cerium is to acts as redox mediator cycling from the oxide form ($CeO_2/Ce_2O_3$), where it acts as electrocatalyst for water reduction to OH⁻ (then converted at the counterelectrode - Pt- to $O_2$) and $Ce^0$ where this species reacts with $CO_2$ to form carbon being oxidized to $CeO_2$. At an optimal potential of -2.0 V (vs Ag/AgCl) a Faradaic efficiency (FE) of about 77% to C ($FE_C$ was indirectly calculated) and 23% to $H_2$ were obtained, without CO formation. Batch tests were only made, without indications of the possibility of continuous operations.

Liu et al. [61] used direct molten carbonate electrolytic splitting of $CO_2$ to carbon nanoplatelets. The chemistry proposed is the following:

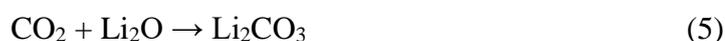

$$CO_2 + Li_2O \rightarrow Li_2CO_3 \quad (5)$$

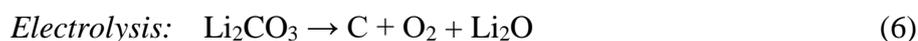

$$\textit{Electrolysis:} \quad Li_2CO_3 \rightarrow C + O_2 + Li_2O \quad (6)$$

Depending on the transition metals added to molten lithium carbonate (metal clusters act as nucleation points for nanocarbon growing), different forms of nanocarbons could be obtained. Liu et al. [61] used zinc to form carbon nano platelets, while in the absence of transition metal nucleating agents carbon nano-onions form [69] or in the presence of other transition metals such as Ni carbon nanotubes (CNT) are formed [70-72] during the molten carbonate electrolysis of $CO_2$. Studies were made in a batch electrochemical cell at about 770°C, using relatively expensive electrodes (Pt Ir foil anode, transition metal coated steel cathodes) without clear indications about the reaction rates, FEs and the possibility of continuous operations, with focus again on the production of the nanocarbon materials rather than on the $CO_2tC$ process. The use of molten lithium carbonate also creates several issues of corrosion as well established for molten carbonates fuel cells [73].

Finally, the work by Wang et al. [74] on the decomposition of $CO_2$ to carbon and oxygen under mild conditions (300°C) over a zinc-modified zeolite should be mentioned, although no further studies on this system were later reported. The formation of carbon is about 900 μmol per g zeolite in 24h of reaction (thus the rate is quite low), and the electrons necessary for the reduction derive from "the electrons delocalized within the $Zn^{2+}$-Y− material", being however, unclear from what they are generated. The reaction appears not to be catalytic.

There are thus some indications in literature of the possibility to realize a continuous $CO_2tC$ process, although relevant data are still missing as well as better indications on how to realize a reliable and effective process to perform this reaction under industrial relevant conditions. It is known that it can be possible to tailor the carbon form obtained depending on various operative parameters and use of substrate catalytic elements, the latter being necessary to operate with enough high reaction rates at low temperatures, where carbon formation is favoured. From this perspective, integrating the use of NTP with some of the approaches described above represents a further possibility to operate at lower temperatures, and eventually avoid the electrolysis step. Furthermore, the objective should be to avoid the energetic step of water electrolysis, present (even if somewhat hidden) in various of the discussed examples.

The following section will first analyse the status in NTP conversion of CO to carbon, to discuss later how to integrate these indications in a possible scheme of plasma-assisted $CO_2tC$ process.

## 3.    Status on plasma-assisted $CO_2$ splitting

The topic of plasma-catalytic $CO_2$ conversion has been widely reviewed in literature [14-18, 29, 37-41]. We will thus recall here only the necessary info to analyse whether integration of NTP within a $CO_2tC$ process scheme is feasible and useful. As indicated in the introduction, a general comment is that only $CO_2$ splitting to CO (rather than to C) or the conversion to oxygenated or hydrocarbons was investigated.

As a general background, may be useful to remember that the behaviour in plasma-assisted $CO_2$ conversion is strongly depending on the type of reactor used for NTP generation. The most common type of NTP reactors investigated for $CO_2$ conversion include dielectric barrier discharges (DBDs), microwave (MW) plasmas, gliding arc (GA) plasmas, atmospheric pressure glow discharges (APGDs), nanosecond-pulsed discharges, plasma vortex and corona and spark discharges. In brief, these different reactors differ in terms of effective contact between the discharging filaments and the gas phase, besides to fluidodynamic determining the secondary processes. Inside the filaments, the conversion and energy efficiency is high, but this is a rather limited fraction of the reactor volume, where either conversion does not occurs or the excited species generated by plasma react with others and can also be quenched. Snoeckx and Bogaerts [17] have discussed in detail the modelling and behaviour in $CO_2$ conversion of the different reactor solutions. DBD, MW, and GA plasma are the most common typology of reactors used. Less common type of plasma reactors include, for example, the Vortex type, which can be considered an advanced reactor design, where by addressing the drawbacks of current gliding arc systems, an improvement of the performances is realized (higher $CO_2$ conversion and energy efficiency) [75]. The performances in all reactors depend drastically on the specific design of the reactor, the space-velocities, the pressure, etc. Trenchev and Bogaerts [75] for a dual-vortex design reported energy efficiencies of around 40% with maximum $CO_2$ conversion around 10% at atmospheric pressure operations. Higher energy efficiencies have been reported for MW and RF reactors, but operating at reduced pressure, where more energy-efficient $CO_2$ conversion can be reached, due to a stronger overpopulation of the vibrational levels [17]. However, reduced pressure operations are not suitable for industrial applications. Higher $CO_2$ conversion, up to over 80-90% could be instead obtained in DBD type reactors, but with low energy efficiencies (less than about 20%).

$CO_2$ splitting under NTP conditions is dominated by i) electron-impact dissociation (forming CO and O atoms and ions), ii) the ionisation process (forming $CO_2$ + ions) and iii) the electron dissociative attachment (forming CO and $O^-$ ions). CO is the main product of $CO_2$ conversion usually reported in the absence of H-donor components. The formation of carbon during NTP $CO_2$ splitting was scarcely reported. However, by spectroscopic analysis of the products formed during inductively heated $CO_2$ plasma Burghaus et al. [76] observed that also C is a major product present, together with CO and O (and other minor species). Thus, both $CO_2$ dissociation to CO + O and CO dissociation to C + O occurs during the plasma contact. The products, and performances, are largely determined from the processes of recombination of the generated species (for example, CO and O) occurring out of the plasma region, and thus largely determined from the reactor geometry and fluidodynamic [77].

Particularly for atmospheric operations, the conversion of $CO_2$ and energy efficiency are typically both below 30%. Thus, the use of this technology alone does not allow an efficient process of $CO_2$ splitting to carbon. However, in principle integrating a CO dissociation catalyst, would result in an improvement of the performances. This is a topic scarcely reported in reviews and articles dealing on $CO_2$ conversion by plasma.

Ashford et al [78] studied the decomposition of $CO_2$ to carbon monoxide and oxygen in a packed bed, dielectric barrier discharge (DBD) reactor at low temperatures (120-130°C) and ambient pressure. They reported a $CO_2$ conversion of 24.5% and energy efficiency of 13.6% using a 5Fe5Ce/γ-$Al_2O_3$ catalyst, but there are no indications on carbon formation. Ray et al. [79] used Ni and Cu oxide supported γ-$Al_2O_3$ and using a packed DBD reactor, reporting a maximum 15.7% conversion of $CO_2$ and energy efficiency < 15%. They observed a carbon balance of about 95%, thus indicating the possible formation of carbon, but in low amounts. Zhang and Harvey (2021) used a DBD packed with $BaTiO_3$ spheres analysing the $CO_2$ decomposition in the presence of up to 50% $O_2$. Even without co-fed oxygen, at atmospheric pressure they observed a CO$_2$ conversion of 15% (declining on increasing O$_2$ concentration) and energy efficiency of 2% also declining on O$_2$ addition. They not observed formation of carbon. Taghvaei et al. [81] studied CO$_2$ decomposition in a DBD plasma reactor packed with a foam coated with various oxides (BaTiO$_3$, TiO$_2$, CeO$_2$, etc.). There is generally an improvement using the foam with respect to the empty reactor, but at the best energy efficiency it is < 9% for a CO$_2$ conversion of around 10%, and <5% for a CO$_2$ conversion of maximum 15% (obtained varying SEI, the specific energy input). They also evaluated the effect of carrier gas type (5% CO$_2$ in He, Ar, N$_2$), which influences the results, but in the range of conversion and energy efficiency indicated before. Taghvaei et al. [81] also not reported the formation of carbon.

Without being exhaustive of the studies on NTP CO$_2$ splitting, the selected recent publications in the area evidence how i) both CO$_2$ conversion and energy efficiency are quite limited (less than 20%) operating at atmospheric pressure without the presence of H-donors (H$_2$O, CH$_4$, etc.) and ii) CO is the product formed, while C formation is either not detected or formed in quite low amounts. Worth to note is that the use of metal grids enhances the CO$_2$ plasma conversion (using a radio frequency-driven plasma reactors) because favour O recombination to form O$_2$, thus reducing relevance of the backreactions [82]. Tests were made at low pressure (100 Pa, about 10$^{-3}$ bar). The CO$_2$ conversion increases from about 30% to 50% in the best case (copper mesh), while energy efficiency was not given. The formation of carbon was not observed. Similar results were also observed by Devida et al. [83].

Yin et al. [85], in their review paper, claim of a new method to get efficient CO$_2$ conversion and high energy efficiency. After analysing plasma and reactor characteristics as well as the mechanisms of CO$_2$ conversion in plasma, they proposed a revised version of the CO$_2$ conversion versus energy efficiency map presented originally by Snoeckx and Bogaerts [17] and updated later also by Bogaerts and Centi [16]. Figure 2 overviews energy efficiency versus CO$_2$ conversion literature results as presented by Yin et al. [85]. Note that data, however, are not homogeneous in terms of pressure of operation and type of plasma reactor (as commented before). Blue start data points are those indicated by Snoeckx and Bogaerts [17]. Orange circle data points those reported by Yin [85,86] using a thermal arc, those of purple square are the results by Bermudez et al. [87] using microwave heating. In both these cases, the Boudouard reaction is addressed by introducing a solid C reactant in the reactor. These data highlighted by Yin et al. [85] to demonstrate the possibility to obtain both superior CO$_2$ conversion and energy efficiency in CO$_2$ splitting, refer to the use of carbon in capturing the O formed in $CO_2$ to $CO + O$ process. In fact, no $O_2$ is emitted and only CO forms. Thus, the Boudouard chemistry is not involved in producing carbon by CO disproportionation, but rather to capture the O generated by $CO_2$ splitting forming two CO molecules (even if this stoichiometry was not demonstrated).

In conclusions, NTP has the potential to be a more energetically favourable process (assuming a high energy efficiency) than the process CO₂tC process involving electrolysis, but the focus of all studies on $CO_2$ splitting was the production of CO, rather than its further splitting to produce C. Data on the latter are fragmentary, and good results with accurate carbon balance are missing. However, in general, the formation of carbon is likely low, even if spectroscopic data show that CO splitting to C + O occurs fast in the discharge zone. Thus, recombination processes are dominant the chemistry of the process.

## 4    A conceptual possibility for CO₂tC process

Having analyzed the status in CO₂tC by catalysis and its limits, and the status in NTP of $CO_2$ splitting, the further question is to analyze the possibility to combine the various discussed elements to propose a different novel approach overcoming current limitations of the previous studies and thus possibly opening new prospects for this challenging reaction.

The starting point is that NTP is effective in both $CO_2$ splitting to $CO + O$, and CO splitting to $C + O$, thus overcoming the limit and energy required to split water by electrolysis to produce on site the H-donor agents ($H^+/e^-$) necessary to convert the O species. However, the effective products determined in $CO_2$ splitting by NTP are largely dominated from the recombination chemistry. To make effective the reaction, it is necessary from one side to catalyse the recombination of two O atoms to an oxygen molecule, then fast removing the latter to avoid its further conversion. The data discussed in section 3 showed that the use of metallic catalysts can be effective in promoting this recombination.

On the other side, it is necessary to identify a chemistry which stabilize the C atoms produced, forming nanocarbon species which result stable and that can be recovered in a continuous manner. The other indication is that the Boudouard chemistry occurs during the reaction, but it is necessary to block in some way (and then recycle) the $CO_2$ produced. We have evidenced in section 2 that nanocarbon type materials can grow under conditions of the Boudouard reaction which is favoured by low temperatures. The electrolysis of the Li-carbonate (eq.s 5 and 6) is an interesting possibility, but occurs at high temperature (about 750°C) where the Boudouard reaction is shifted to the formation of CO, besides to the problems of corrosion and management of the process under these operative conditions. On the other hand, it was shown that by using CaO instead of $Li_2O$ (with great advantages in terms of costs and reduced corrosion issues) there is a clear advantage in the Boudouard reaction by feeding CO rather than $CO_2$ [88]:

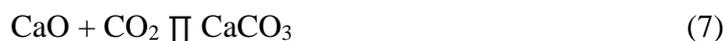

$$CaO + CO_2 \rightleftarrows CaCO_3 \qquad (7)$$

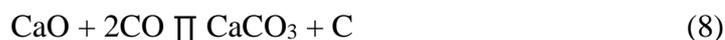

$$CaO + 2CO \rightleftarrows CaCO_3 + C \qquad (8)$$

Rout et al. [88] calculate that the equilibrium partial pressure of CO for eq. (8) is much lower than that for the Boudouard reaction at all the temperatures studied, indicating a sorption enhanced effect. Moreover, for the eq. (8) at 400 °C, the CO and $CO_2$ partial pressure values are close to zero (Figure 3), while between 400 and 450 °C only CO is present. Huang et al. [89] showed that (Li–Na–K)₂CO₃ molten salts coated on CaO particles greatly improve the adsorption capacity and kinetics (by a factor up to about 4) of CaO with respect to $CO_2$ adsorption even at low temperatures (400-500°C). This system will also have enough conductivity to act as electrolyte to apply a potential for the electrolysis of Ca-carbonate at temperatures in the 400-500°C:

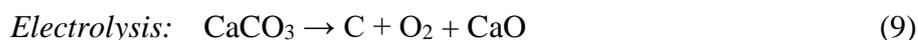

$$\textit{Electrolysis:} \quad CaCO_3 \rightarrow C + O_2 + CaO \qquad (9)$$

The C formed will react with the other C species generated either by the Boudouard reaction and those generated in the gas-phase by NTP CO splitting. A transition metal catalyst dissolved in the molten salt will favour the growth of the nanocarbon species. The NTP would be necessary to promote the low temperature (<600°C) chemistry and kinetics of the process.

We can combine all the discussed elements in the schematic illustration of the $CO_2tC$ device presented in Figure 4. An array of plasma jet units [90] to generate the plasma is used. At a short distance a metallic foam (Cu or other metals) is placed with the function to catalyse the recombination of the generated O species (from $CO_2$ and Co splitting) to produce $O_2$. The plasma effluents then enter in contact with a molten salt (maintained at a temperature of 400-500°C) containing CaO nanoparticles (NPs) and transition metal NPs (Zn or other metals) which catalyse the aggregation of C to form nanocarbon species. The application of a potential will convert the Ca-carbonate species formed generating C and $O_2$ (eq. 9).

This it is a conceptual scheme, aimed to combine the different info discussed in this paper to look at a possible alternative scheme of a device overcoming some of the limitations in the actual proposed process schemes for $CO_2tC$. Possible alternatives are feasible, and the proof of this proposed device must be demonstrated. The aim here, however, is not to indicate that this scheme works and is superior to other proposed solutions. As discussed in the previous sections, there is still not a valid and reliable solution for this challenging reaction of splitting $CO_2$ to C and $O_2$. On the other hand, still very limited indications are present on this reaction in literature. The aim here is thus to stimulate the investigation of this reaction, with the scheme of device proposed in Figure 4 as a possibility to use in an unconventional way the literature indications, but also to encourage to consider alternatives and improvements.

## 5    Conclusions

This concept review paper analysed the possibility to develop a technical solution for the challenging reaction of $CO_2$ splitting to carbon and $O_2$ ($CO_2$tC). This is an important possibility for some specific applications (for example, recovery of $O_2$ from $CO_2$ in space missions), but also as alternative sustainable solution for the storage of $CO_2$ or its conversion to chemicals. There are quite limited literature indications that were reviewed, with an analysis also of the limits. Not-thermal plasma, in combination with catalysis, is one of the possibilities to promote this reaction, but the current studies on $CO_2$ splitting are limited mainly to the formation of CO. By combining data on the status of current studies on NTP $CO_2$ splitting with those on $CO_2$tC it is possible to propose the novel scheme of device illustrated in Figure 4, but which aim is to stimulate creativity of the readers to consider alternative possibilities for this challenging reaction, rather than to demonstrate feasibility of the proposed technology.

## Acknowledgements

This project has received funding from the European Research Council (ERC) under the European Union's Horizon 2020 research and innovation programme (grant agreement No 810182). The authors would thank also the PRIN2017 support by the project $CO_2$ ONLY (grant number 2017WR2LRS).

## Author's contributions
G. Centi and S. Perathoner equally contributed to prepare and revise the manuscript.

**FIGURE CAPTIONS**

**Figure 1** Schematic representation of the ETHL process for $CO_2$ splitting to carbon and oxygen. Elaborated from the original reported by Luc et al. [42].

**Figure 2** Summary of results of $CO_2$ conversion versus energy efficiency in literature data on NTP $CO_2$ splitting. Data refer to different type of plasma reactors, operation conditions and pressure. Blue start data points are those indicated by Snoeckx and Bogaerts [17]; orange circle data points those reported by Yin [85,86] using a thermal arc and a solid C target, those of purple square are the results by Bermudez et al. [87] using the Boudouard reaction driven by microwave heating of a solid C reactant in a $CO_2$ flow. Elaborated from the original reported by Yin et al. [85].

**Figure 3** Equilibrium partial pressure of CO vs. temperature for the Boudouard reaction (eq. 3), and equilibrium partial pressures of CO and $CO_2$ vs. temperature for the sorption enhanced Boudouard reaction (eq. 8). Elaborated from the original reported by Rout et al. [88].

**Figure 4** Proposed simplified scheme for a device for $CO_2tC$ process, based on the combination of non-thermal plasma and catalysis.

**FIGURES**

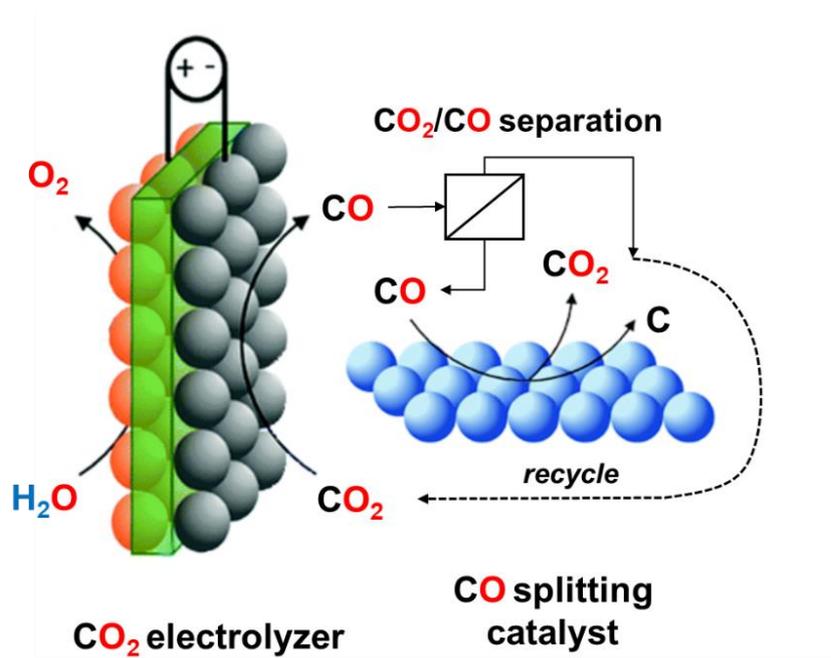

FIG. 1

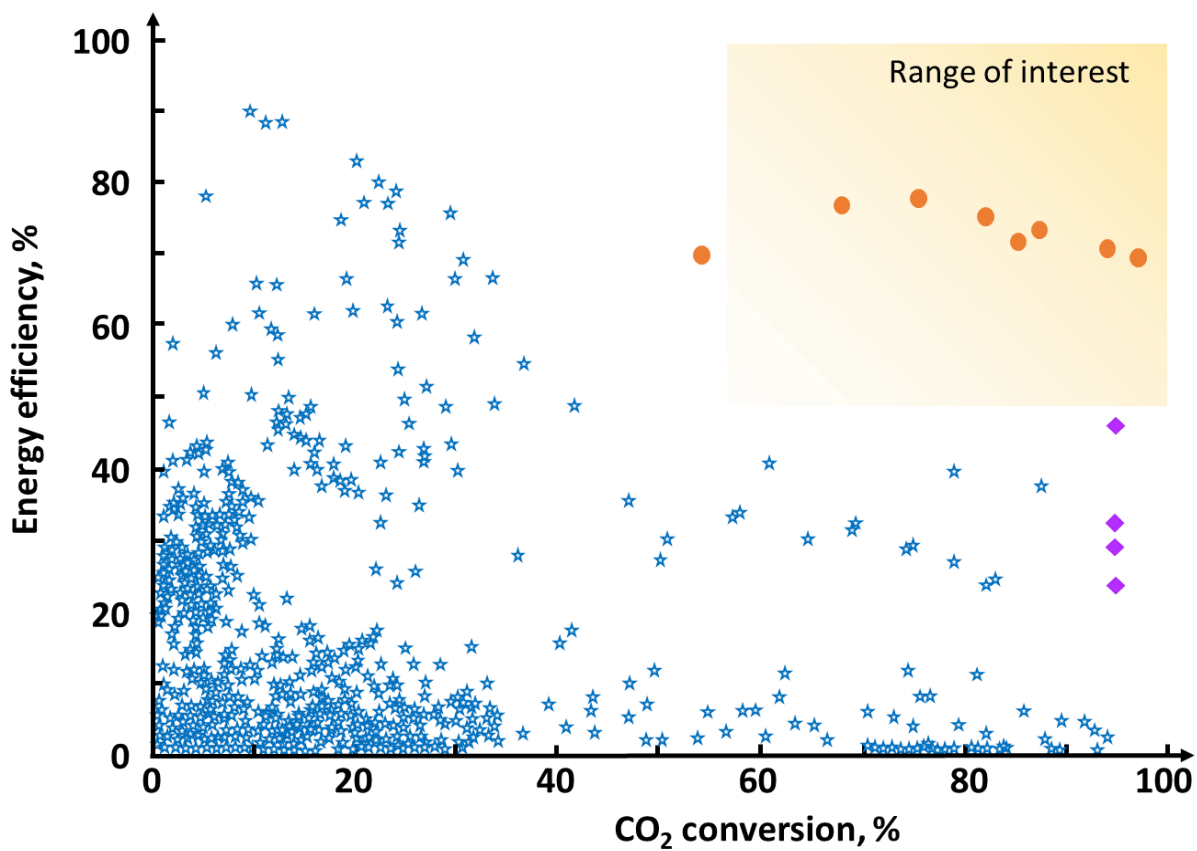

FIG 2

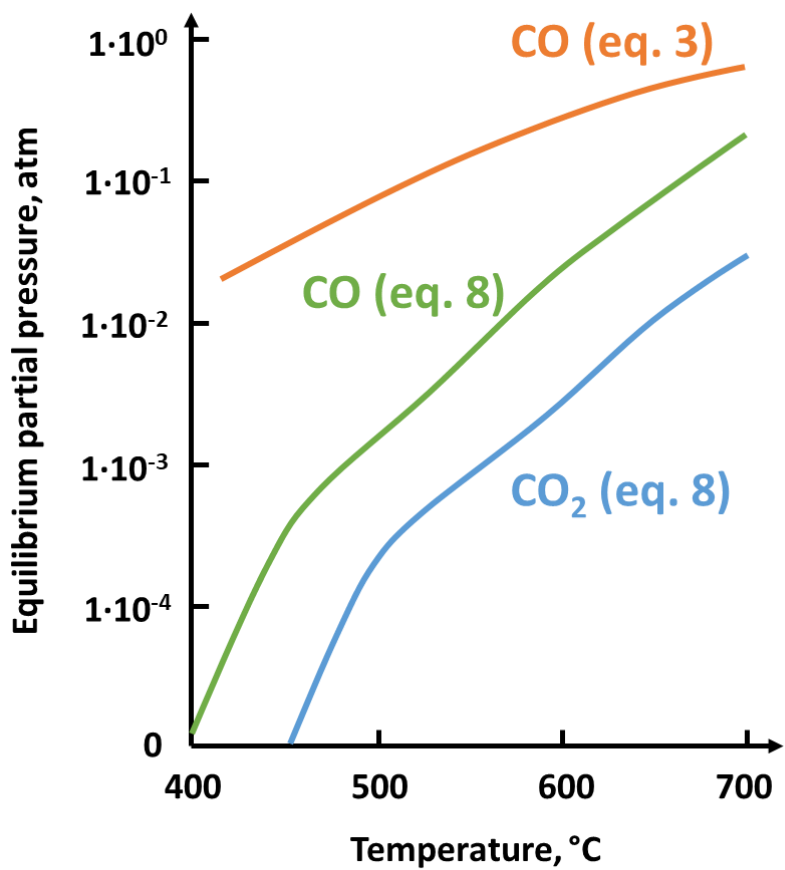

FIG. 3

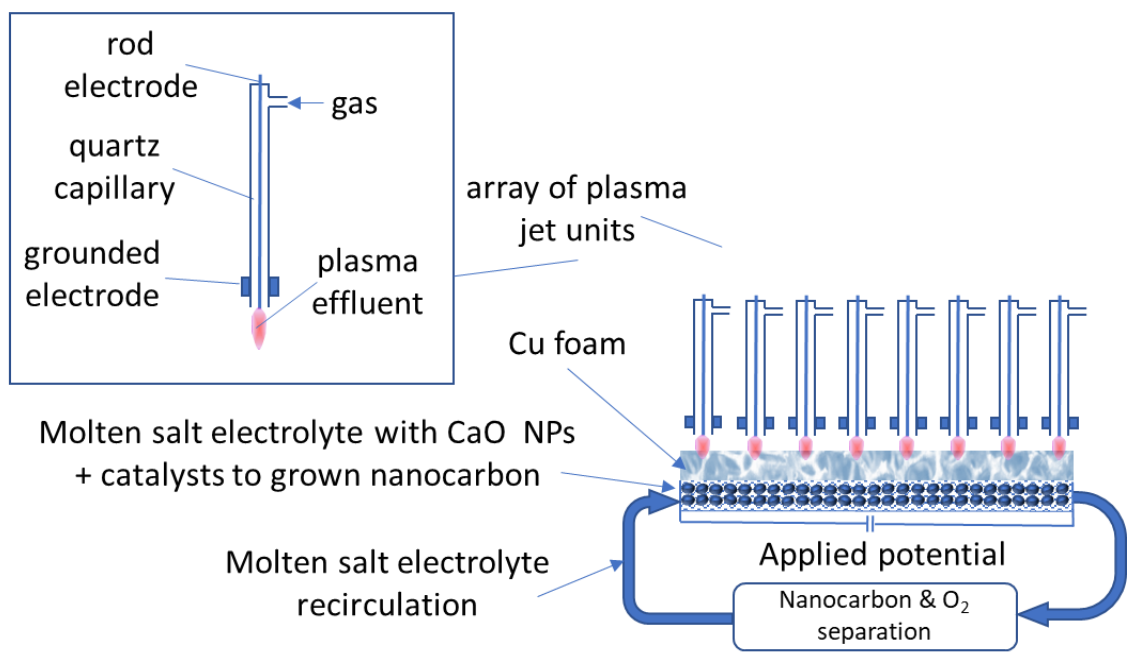

FIG. 4